\documentclass[10pt]{iopart}
\usepackage{graphicx}

\begin{document}

\title[Effect of uranium deficiency in unconventional superconductor UTe$_2$]{Effect of uranium deficiency on normal and superconducting properties in unconventional superconductor UTe$_2$}

\author{Y. Haga, P. Opletal, Y. Tokiwa, E. Yamamoto, Y. Tokunaga, S. Kambe and H. Sakai}

\address{Advanced Science Research Center, Japan Atomic Energy Agency}
\ead{haga.yoshinori@jaea.go.jp}
\vspace{10pt}
\begin{indented}
\item[]\today
\end{indented}

\begin{abstract}
Single crystals of the unconventional superconductor UTe$_2$ have been grown in various conditions which result in different superconducting transition temperature as well as normal state properties.
Stoichiometry of the samples has been characterized by the single-crystal X-ray crystallography and electron microprobe analyses.
Superconducting samples are nearly stoichiometric within an experimental error of about 1 \%, while non-superconducting sample significantly deviates from the ideal composition. 
The superconducting UTe$_2$ showed that the large density of states was partially gapped in the normal state, while the non-superconducting sample is characterized by the relatively large electronic specific heat as reported previously.

\end{abstract}

%
%
%
%
\ioptwocol

\section{Introduction}



Uranium ditelluride UTe$_2$ was recently reported to show unconventional superconductivity \cite{Ran2019}. 
It crystallizes in the orthorhombic $Immm$ structure of its own type \cite{Haneveld1970,Beck1988}.
Unlike other two uranium dichalcogenides $\beta$-US$_2$\cite{Suski1972a} and $\beta$-USe$_2$ \cite{Shlyk1995a} with the different crystal structure, UTe$_2$ shows metallic conductivity down to low temperatures.\cite{Shlyk1999} Uranium 5$f$ electrons carry magnetic moment roughly corresponding to uranium paramagnetic effective moment.\cite{Suski1973}  
Single crystal study showed that the paramagnetic behavior is anisotropic with the easy axis along the $a$-axis.  
The paramagnetic susceptibility along the easy axis monotonically increases with decreasing temperature down to 2 K without showing a magnetic phase transition.\cite{Ikeda2006} 
The relatively large electronic specific heat coefficient 150 mJ/K$^2$mol \cite{Ikeda2006} characterizes UTe$_2$ as a moderately heavy fermion system. 
Superconducting transition was discovered below 2 K for a high quality single crystal with lower residual resistivity than the previous single crystal.\cite{Ran2019}  The superconducting transition accompanies large specific heat anomaly arising from the enhanced electronic specific heat, directly demonstrating that the superconductivity is realized by conduction bands with 5$f$ characters. 
In fact, considerable amount of 5$f$ contribution is observed in photoemission spectra \cite{Fujimori2019,Miao2020}. The magnetic excitation spectra characterized by a low-dimensional magnetic fluctuation\cite{Knafo2021} is also influenced across the  superconducting transition\cite{Duan2021,Raymond2021}, indicative of a magnetically mediated pairing interaction.

The extremely anisotropic superconducting upper critical field $H_{\rm c2}$ and, in particular, unusual increase of $H_{\rm c2}$ for the field along the $b$-axis has similarity to the phenomenon observed in ferromagnetic superconductors\cite{Sheikin2001,Levy2005a,Huy2008,Aoki2009a}, making UTe$_2$  a possible candidate for a superconductor with a nearly ferromagnetic electronic state and a spin-triplet pairing.\cite{Ran2019} Unconventional nature of superconductivity such as time-reversal symmetry breaking and multiple superconducting transitions is actively discussed\cite{Hayes2021}. For the latter issue, particularly, it is important to rule out the possibility of multiple transitions arising from the sample dependence. In ref. \cite{Thomas2021} it was demonstrated that superconducting regions with different $T_{\rm c}$ can be found in a bulk sample. 

Proximity to the magnetically ordered phase is demonstrated by high-field or high-pressure investigations. Application of the strong magnetic field along the magnetically hard axis $b$ induces a phase transition to a ferromagnetic-like state through a first-order transition. 
The field-induced reentrant superconductivity for the field between $b$ and $c$ axis seems to coexist with the polarized moments\cite{Ran2019a}. 
On the other hand, possible magnetic phase transition appears under pressure above the relatively low critical pressure where the magnetic anisotropy drastically changes as a function of pressure. \cite{Braithwaite2019,Li2021}

As inferred from the absence of superconductivity in single crystals used for earlier studies, relatively small lattice imperfection has strong impact on physical properties. Even in superconducting samples, superconducting transition temperature varies from about 1.5 K up to 2 K\cite{Hayes2021,Cairns2020,Rosa2021}. Previous study proposed that the emergence of superconductivity would be caused by slight Te-deficiency \cite{Cairns2020}. Yet, the relationship between such non-stoichiometry and physical properties has not been clarified.

In this paper we investigate the crystal structure and physical properties of UTe$_2$ with different preparation conditions. In particular, the sample dependent crystallographic structure in relation with the physical properties is discussed.

\section{Experimental}

Single crystal samples of UTe$_2$ were prepared with chemical vapor transport (CVT) method using iodine as the transport agent.
Starting uranium metal, tellurium and iodine were introduced in an evacuated quartz tube.
Temperature gradient was applied by two individually controlled  heaters.

Single crystal X-ray diffraction was measured with a graphite monochromated Mo K$_\alpha$ radiation. The scattered X-ray beam was recorded on an imaging plate detector (R-AXIS RAPID, Rigaku). Absorption correction with the empirical method was applied prior to the structural solution. Structural solution and crystallographic parameters fitting were performed by SHELX program\cite{Sheldrick2015}.
Electron-probe microanalysis was performed by wavelength-dispersive spectrometers installed in a scanning electron microscope (JXA-8900, JEOL).
Electrical resistivity was measured using ac four-probe method in a closed-cycle refrigerator. Specific heat was measured using the heat relaxation method with a home-built calorimeter installed in a $^3$He cryostat.

\section{Results}
\subsection{Single Crystal Growth}

Table~\ref{t1} summarizes the characterization of representative samples grown from different conditions. Among them, the sample NS was grown following the condition described in ref. \cite{Ikeda2006}. The resistivity and specific heat measurements showed absence of superconductivity down to 0.45 K. On the other hand,  SC1 samples grown following the Ran's prescription \cite{Ran2019}, where tellurium deficient starting composition is employed, showed superconductivity around 1.6 K. In the SC2 samples, we kept the starting composition the same but applied lower temperature condition. The resulting crystals showed fairly high superconducting transition temperatures. This demonstrates that the starting composition affect $T_{\rm c}$ and residual resistivity ratio (RRR) rather than the growth temperature. Higher $T_{\rm c}$ samples with $T_{\rm c}$ up to 2 K were recently reported from much lower growth temperatures\cite{Rosa2021}. The RRR for the superconducting samples are always higher than that of non-superconducting sample, indicating that the non-superconducting sample shows more electron scattering due to impurities or defects.

\begin{table}
\caption{\label{t1}Summary of UTe$_2$ single crystals studied in this paper. All the samples were prepared from iodine vapor transport method but with different conditions. U:Te means starting composition. $T_{\rm c }$ is defined as the onset of resistivity drop. RRR means residual resistivity ratio.}
\footnotesize
\begin{tabular}{@{}lllll}
\br
Batch &  U:Te & Temperature  ($^\circ$C) & Resistive $T_{\rm c}$ & RRR \\
\mr
NS & 1:2 & 950/850 &  $< 0.45$ & 1.7 \\
SC1 & 1:1.8 & 1050/1000 & 1.52-1.63 & 15-33 \\
SC2 & 1:1.8 & 950/850 & 1.75-1.90 & 20 \\

\br
\end{tabular}\\

\end{table}
\normalsize

\subsection{Crystallographic characterization}

\begin{table}
\caption{\label{t2}Typical values for crystallographic parameters for superconducting and non-superconducting UTe$_2$ samples. Experimental uncertainty in parenthesis includes deviations obtained for samples belong either to NS and SC1/SC2 type. }
\footnotesize
\begin{tabular}{@{}rrr}
\br
 &  NS samples  & SC1/SC2 samples \\
\mr
\multicolumn{1}{l}{lattice parameters  (${\rm \AA}$) } & &\\
$a$ &4.1600(2) & 4.1618(5)\\
$b$ & 6.1219(4) & 6.1355(7) \\
$c$ & 13.9476(9) & 13.9698(13) \\
\mr
\multicolumn{1}{l}{fractional coordinates) } & &\\
U $z$ & 0.13545(5) & 0.13520(4) \\
Te1 $z$ &  0.29755(5) & 0.29780(10) \\
Te2 $y$ & 0.24895(5) & 0.24910(3) \\
\mr
\multicolumn{1}{l}{$B_{\rm eq}$ $^a$ } & &\\

U & 0.687 & 0.513 \\
Te1 & 0.562 & 0.576 \\
Te2 & 0.515 & 0.529 \\
\mr
\multicolumn{1}{l}{reliability factors $^b$ } & & \\
$R_1$ & 0.0222 & 0.0309 \\
$wR_2$ & 0.0564 & 0.0586 \\

\br
\end{tabular}\\
$^a$ Equivalent isotropic atomic displacement parameter is defined as 
$B_{\rm eq} = 8\pi^2/3 [ U_{11}(aa^*)^2 + U_{22}(bb^*)^2 + U_{33}(cc^*)^2 ]$ where $U_{ij}$ is experimentally obtained anisotropic atomic displacement tensor. $^b$ $R_1$ and $wR_2$ are unweighted and weighted agreement factors, respectively.

\end{table}
\normalsize

\begin{figure}
\begin{center}

\includegraphics[width=7.5cm]{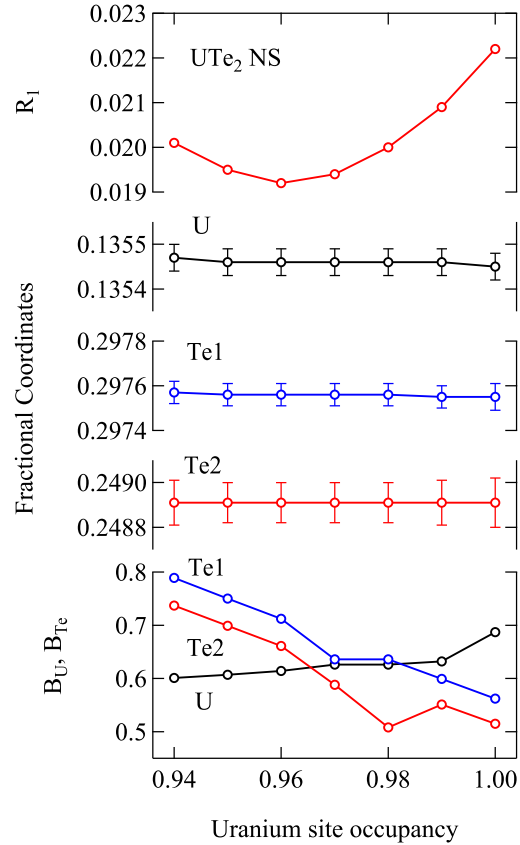}
\end{center}
\caption{Agreement factor, fractional coordinates and equivalent isotropic atomic displacement parameters as a function of the occupancy parameter at the uranium site.}
\label{fit_XRD}

\end{figure}

Table~\ref{t2} shows the crystallographic parameters determined by single-crystal X-ray diffraction.
Fractional coordinates, anisotropic atomic displacement parameters and secondary extinction effect were taken into account during the fitting.
The lattice parameters of sample NS are slightly smaller than SC1 and SC2, while there are no significant differences in the fractional coordinates. 
We note that the atomic displacement parameter for uranium site in NS is fairly large. It is even larger than that of tellurium sites.
Such a large atomic displacement parameter is indicative of a large spread of spatial distribution or low atomic concentration at the corresponding site.
To clarify this we attempt to fit the data by changing the uranium occupancy. As shown in Fig. \ref{fit_XRD} fractional coordinates are very insensitive to the uranium occupancy. On the other hand, the reliability factor $R_1$ shows a minimum when the occupancy is around 0.96. At this occupancy, the atomic displacement parameter of uranium is smaller than the two tellurium sites, as seen in most of conventional metallic compounds. The similar analysis for the samples SC1 and SC2 converges around full uranium occupancy. Uranium deficiency was found also in uranium monotelluride UTe. See Appendix for details.

The occupancy 0.962(2) of the uranium ($Z$ = 92) site is equivalent to the existence of an atom with $Z$ = 88.3. 
There are two possibilities to realize such a situation. One is that $\sim 4$\% of uranium atoms are missing, and the other is that the extra tellurium ($Z$ = 52) partially substitute the uranium site. In the latter case, the resulting composition would be (U$_{\rm 0.91}$Te$_{0.09}$)Te$_2$ = U$_{0.91}$Te$_{2.09}$ (U$_{0.87}$Te$_2$).

To verify this, the chemical composition of the samples is examined with the electron-probe microanalysis. There is a clear difference in the intensity ratio between signals from uranium and tellurium. By using the intensity ratio for sample SC2 as an internal standard for stoichiometric UTe$_2$, the composition of the sample NS is estimated as U$_{0.96\pm0.01}$Te$_2$. It is therefore likely that uranium site is not substituted by Te but rather left vacant to satisfy average 96 \% occupancy. The possibility of the excess atoms at interstitial positions is examined by constructing an electron density map from the X-ray diffraction data by maximum entropy method \cite{Momma2013}. The resulting electron density did not show peaks higher than 1 e/$\rm \AA^3$ at interstitial positions. Therefore this possibility can be excluded.


In the previous study of U-Te binary phase diagram, the stoichiometric UTe$_2$ was reported to appear by peritectic reaction with Te-rich liquid and U$_3$Te$_5$ phases at 1180$^\circ$C, and then non-stoichiometry of UTe$_{2-x}$ was reported to range from $x=0$ to $x=0.13$ \cite{Boehme1992}, although the tetragonal UTe$_{1.87}$ phase was questionable \cite{KleinHaneveld1964}. Indeed, the Te-flux method using Te-rich composition yields single crystals of UTe$_2$ when temperature was gradually lowered from 1180 $^\circ$C \cite{Ran2021a}, so this appears to be consistent with the phase diagram. However, the crystals grown by the Te-flux method \cite{Ran2021a,Aoki2019b} do not show  superconductivity or show $T_{\rm c}$ less than $\sim$1.6 K, which may correspond to our NS sample. It may suggest that the composition on the peritectic line at 1180 $^\circ$C would slightly deviate from UTe$_2$, {\it i.e.}, U$_{1-x}$Te$_2$. Of course, in the case of CVT method, it is not in the chemical phase equilibrium, the phase diagram cannot be directly applicable.
So far, for the synthesis of superconducting samples by CVT, a necessary condition is considered to start from a U-rich composition for UTe$_2$, {\it i.e.}, U:Te=1:1.5, or 1:1.8. This requirement is, at least, consistent with the off-stoichiometric tendency near UTe$_2$ and our experimental result that the U-deficient crystals do not show superconductivity. 

The sample dependence on superconductivity and chemical composition is extensively studied in ref. \cite{Cairns2020} and a significant difference in composition ratio Te/U is reported between superconducting and non-superconducting samples, where the non-superconducting sample shows systematically larger Te/U ratio. Qualitatively, their observation is consistent with the present EPMA result. In addition, we showed crystallographically that Te/U ratio is not due to the presence of excess Te, but due rather to uranium deficiency.

The deviation from the stoichiometry should also be reflected in the crystallographic parameters. 
As summarized in Table II, lattice parameters of NS1 sample is slightly smaller than those of SC samples. The difference in the unit cell volume is about 1.2 ${\mathrm \AA}^3$. 
Assuming that uranium atoms occupy roughly 1/3 of unit cell volume, a 4 \% deficiency of uranium would result in 1.3 \% of volume reduction if the vacancy is filled by other atoms.
The fairly small experimental volume reduction 0.5 \% suggests that the uranium deficient site is left vacant rather than being substituted by other atoms.

The difference in the fractional coordinates is extremely small between SC and NS samples, except for the $z$-coordinate of the uranium site.
 U$_z$ = 0.1354(1) for NS samples is slightly larger than that of SC samples 0.13520(4). The resulting U-U distance 3.78 is, however, almost sample independent. Overall crystal structure seems to be maintained primarily by the tellurium lattice.
 
 \subsection{Normal and Superconducting properties}
 
\begin{figure}
\begin{center}

\includegraphics[width=8.5cm]{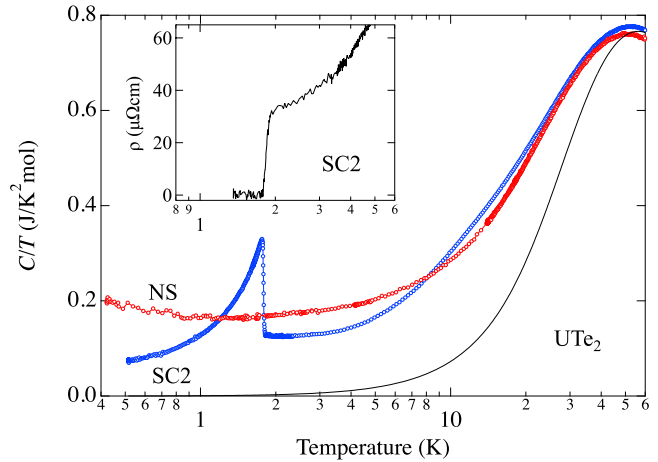}
\end{center}
\caption{Temperature dependence of specific heat divided by temperature for UTe$_2$ NS and SC samples. Inset shows electrical resistivity for SC2 sample near the superconducting transition temperature.}
\label{resis_cot}

\end{figure}

Temperature dependences of electrical resistivity and specific heat measured on the same piece of single crystal SC2 are shown in Fig.~\ref{resis_cot}. Both resistivity and specific heat show a sharp transition at 1.8 K. The onset of the resistive transition begins at slightly higher temperature 1.93 K than the onset of specific heat anomaly 1.81 K, indicating that superconductivity is not occurring in entire sample in this temperature interval. 
Below 1.81 K, specific heat drastically increases, shows a sharp peak at 1.76 K and then decreases. With decreasing temperature, specific heat monotonically decreases down to 0.5 K. Unlike previous works reporting multiple superconducting transition, only a single and sharp superconducting transition is detected in this temperature range. Application of magnetic field along the $a$-direction (not shown) monotonically shifts the peak to the lower temperature side without splitting.

Specific heat below 1.3 K is approximately reproduced by $C/T = \gamma_{\rm N} + \beta T^2$ with $\gamma_{\rm N}$ = 0.047 J/K$^2$mol, where $\gamma_{\rm N}$ is considered as the residual electronic specific heat in the superconducting state. It is now known that the $\gamma_{\rm N}$ correlates with $T_{\rm sc}$.  The most recent study on the sample with higher $T_{\rm sc}$ = 2 K shows less $\gamma_{\rm N}$. It is expected that the highest $T_{\rm sc}$ sample would have significantly small $\gamma_{\rm N}$ term. \cite{Rosa2021,Aoki2021}

Specific heat is sample dependent both in low temperature and high temperature regions. 
In order to minimize experimental uncertainty in obtaining the absolute values, samples with the similar mass 10 mg were measured, where the molar specific heat is calculated using the nominal UTe$_2$ composition. 
At high temperature around 60 K, the electronic contribution to specific heat is expected to be less significant. The difference of specific heat can be regarded primarily as the difference of the lattice contribution. The smaller values for NS sample is indicative of a higher Debye temperature, most likely in agreement with the smaller lattice parameters for NS sample. With decreasing temperature $C/T$ value shows a maximum around 50 K, then decreases with decreasing temperature. $C/T$ of SC sample is larger at high temperatures, but crosses at 8 K. Below this temperature $C/T$ of SC sample is smaller. The difference becomes more significant with lowering temperature down to $T_{\rm sc}$ of SC sample. Note that, by taking into account the uranium deficiency in NS sample, uranium contribution to $C/T$ is even larger.

Between 8 and 2 K, $C/T$ for NS sample approximately follows $\gamma + \beta T^\delta$ with $\gamma = 158~$mJ/K$^2$mol, $\beta$ = 0.0039 J/K$^{2.5}$mol and $\delta = 1.5$. A gradual increase of $C/T$  is observed below 1 K. The origin of this anomaly is not clear at present. In SC sample, on the other hand, the normal state $C/T$ cannot be reproduced by the same manner; $\delta$ is strongly temperature dependent. The large electronic contribution around 14 K, which is also reported by Willa {\it et al.}, is prominent only for SC2 but is absent in NS sample.\cite{Willa2021}

\begin{figure}
\begin{center}

\includegraphics[width=8.5cm]{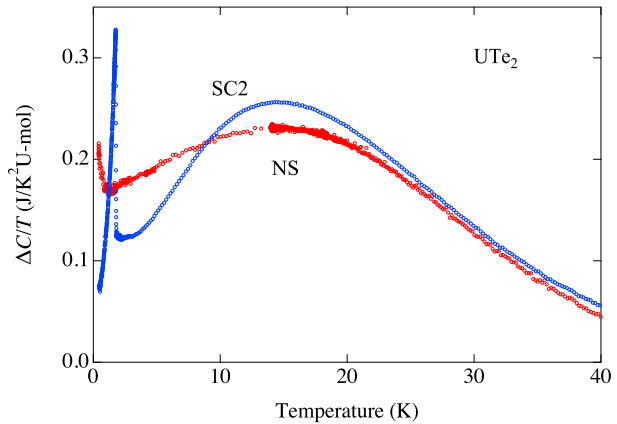}
\end{center}
\caption{Temperature dependence of electronic specific heat for NS and SC2 samples.}
\label{delta_CoT_Umol}

\end{figure}

In order to estimate electronic contribution to specific heat, phonon contribution is approximated by Debye model with $T_{\rm D}$ = 200 K as shown by black line. The resulting electronic specific heat $\Delta C/T$ was obtained by subtracting Debye model from the experimental data (Figure \ref{delta_CoT_Umol}). Here, the specific heat for NS sample is corrected for uranium deficiency.

In the SC sample, $\Delta C/T$  just above $T_{\rm c}$ is characterized by a heavy fermion behavior with Sommerfeld coefficient 120 mJ/K$^2$mol. Assuming that the heavy fermion state arises from Kondo-type interaction, characteristic energy scale would be of the order of 100 K. The corresponding specific heat associated with the heavy fermion formation would then follow Coqblin-Schrieffer model \cite{Coqblin1969}; the heavy fermion contribution would gradually decrease with increasing temperature. 
In experiment, however, a convex feature around 14 K is seen with increasing temperature. Subtracting the heavy fermion contribution mentioned above, remaining $\Delta C/T$ would look like a Schottky-like broad peak as also observed by Willa {\it et al}. \cite{Willa2021}.

In the NS sample, on the other hand, this feature is broadened; NS sample shows a lower maximum around 14 K, and $\Delta C/T$ decreases slowly with temperature and shows a larger electronic specific heat coefficient at 2 K. $\Delta C_{\rm NS}$ and $\Delta C_{\rm SC}$ cross at 8.2 K. Interestingly, the area defined by the two curves are approximately the same below and above this temperature. It means that entropy difference between 2 K and  about 30 K is almost the same for both samples. The broad peak at 14 K in SC sample suggests the existence of a kind of energy gap with an excitation gap about 40-50 K. It seems that this gap does not open in the NS sample. The excess Sommerfeld term in NS sample may be accounted for by failing to open a gap in the magnetic/electronic excitation.

Recent inelastic neutron scattering (INS) experiments suggest a development of one dimensional ladder-type magnetic correlation characterized by a ferromagnetic coupling between nearest neighbor uranium atoms \cite{Knafo2021}. The characteristic temperature scale for this correlation is estimated as $T^*_1 \sim 14$ K in agreement with the present observation. The present excitation 'gap' is not completely consistent with the INS observation which suggests rather quasielastic response. However it is not easy in this case to distinguish quasielastic to inelastic responses because UTe$_2$ has enhanced density of states arising from heavy electron and consequently a quasi elastic scattering is expected.

The absence of the Schottky anomaly in the uranium-deficient NS sample is naturally understood by the breaking of the magnetic correlation between nearest uranium atoms. Note that 4 \% of deficiency corresponds to one broken ferromagnetic neighbor in 13 ladder rungs ($\sim 50~{\rm \AA}$). Such defects would easily hinders the development of one-dimensional magnetic correlation.

\section{Summary}

We examined details of the sample dependence of UTe$_2$. Both superconducting and non-superconducting samples crystallize in the orthorhombic crystal structure, except for the fact that NS sample shows significant uranium deficiency. 
Correspondingly specific heat data measured on both samples differ significantly even in the normal state between 2 and  30 K. The Schottky-like anomaly with a maximum at 14 K is extremely broadened in NS sample. This suggests that magnetic excitation spectrum is completely different in both samples. The low dimensional nature of the ferromagnetic correlation observed in neutron scattering experiments would lead to the sensitivity on the uranium deficiency. 

\section*{Acknowledgements}
We thank D. Aoki, Y. Homma and K. Kaneko for discussions. This work is financially supported by JSPS KAKENHI Grant Numbers 16KK0106, 19K03726,  20H00130 and 20KK0061.

\clearpage
\appendix

\section{Single crystal X-ray diffraction analysis for uranium site occupancy}
\begin{figure}
\begin{center}

\includegraphics[width=8.5cm]{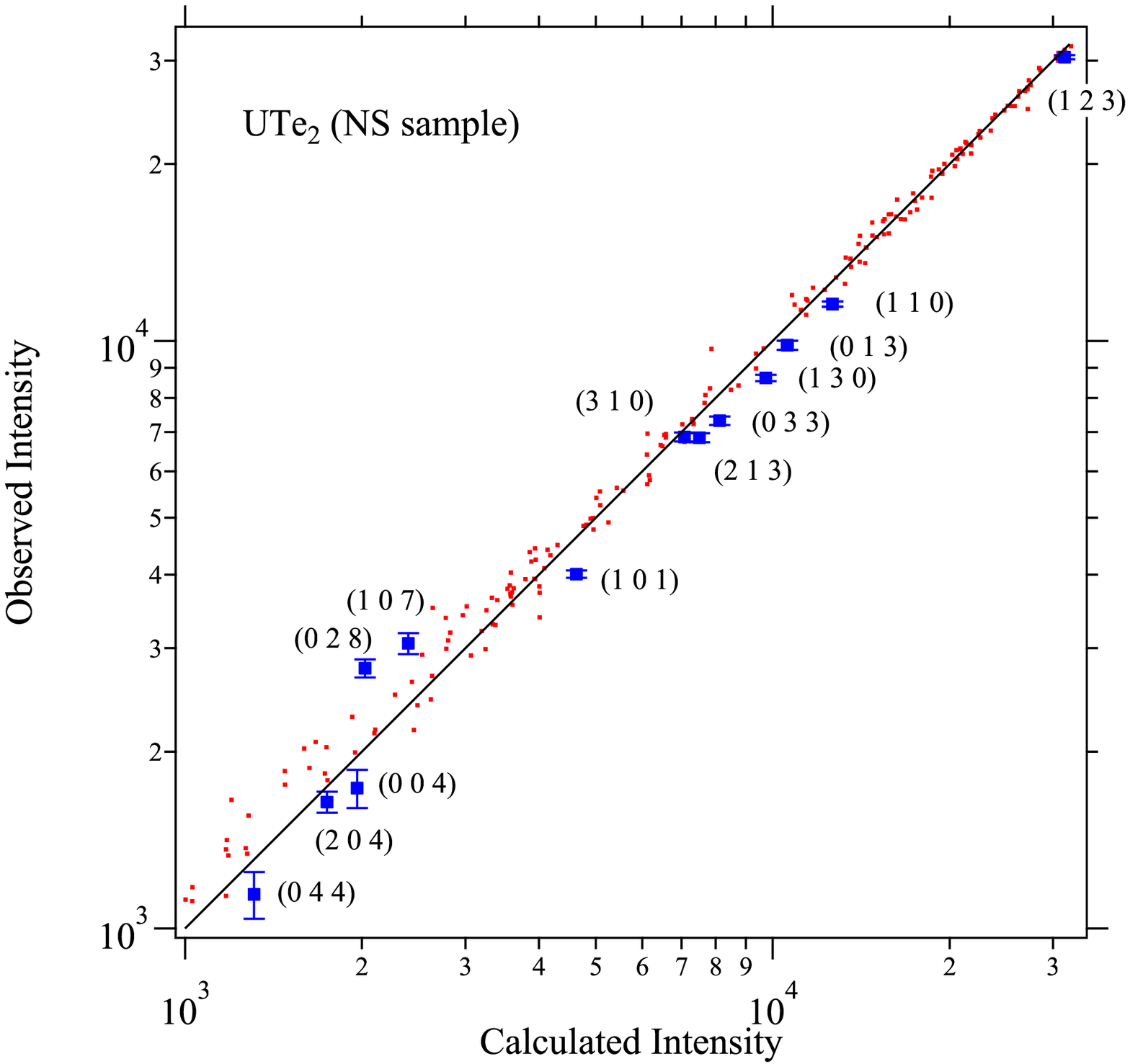}
\end{center}
\caption{XRD intensity for NS sample is plotted against calculated intensity. Uranium site occupancy is fixed to 1.}
\label{XRD_NS}

\end{figure}

\begin{figure}
\begin{center}

\includegraphics[width=8.5cm]{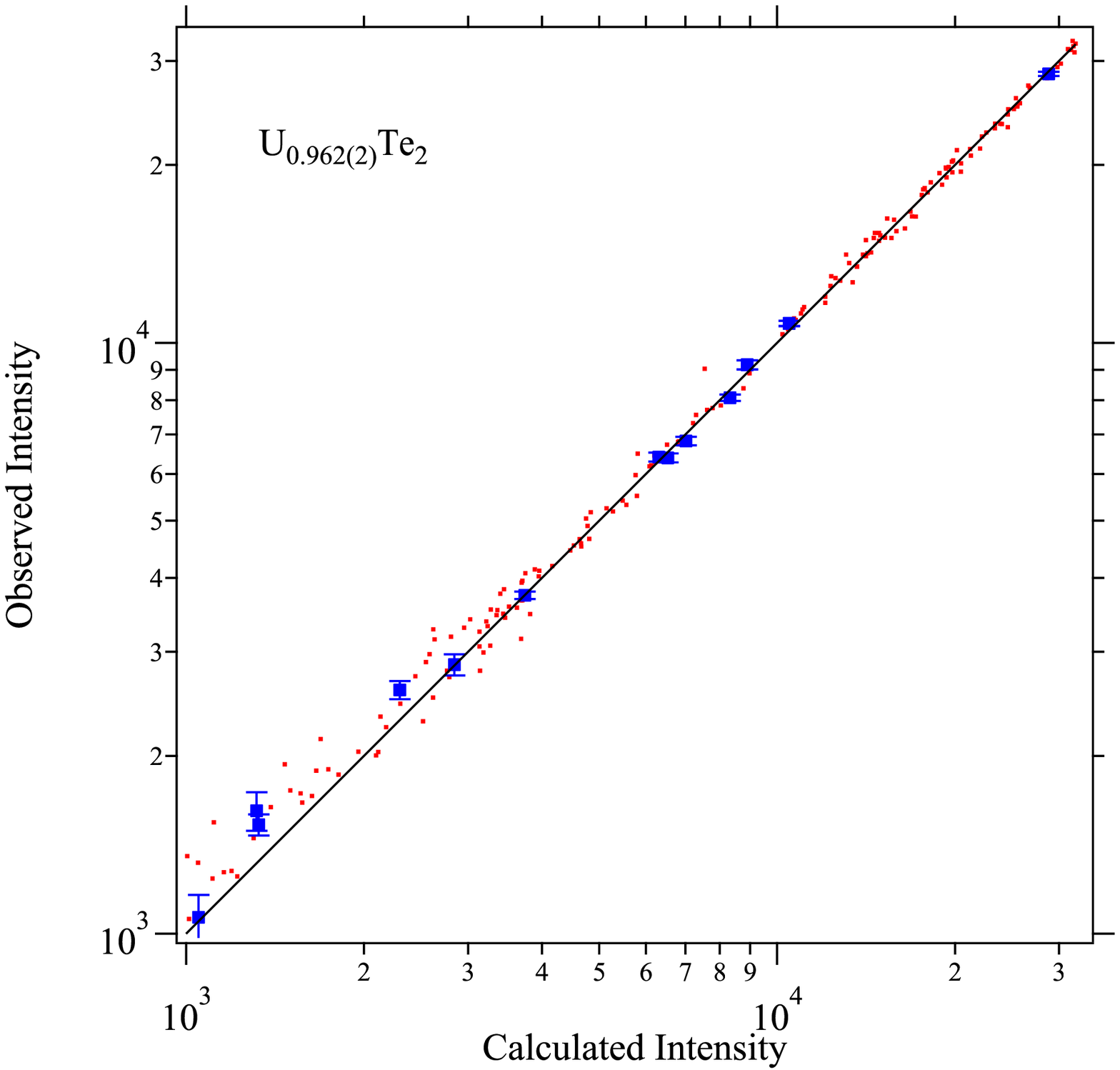}
\end{center}
\caption{XRD intensity for NS sample is plotted against calculated intensity. Uranium site occupancy is fixed to 0.962.}
\label{XRD_NS_U0962}

\end{figure}

\begin{figure}
\begin{center}

\includegraphics[width=8.5cm]{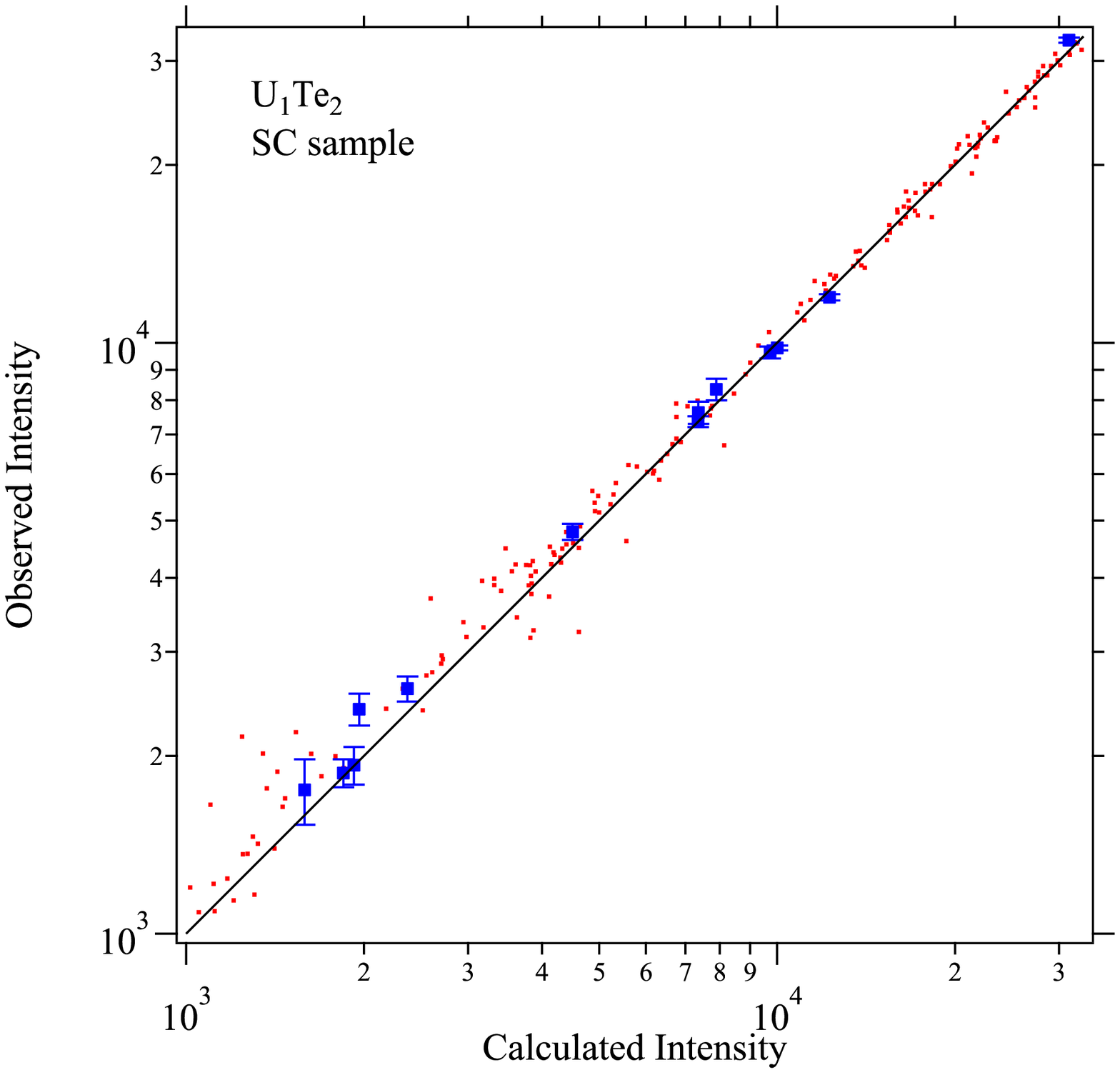}
\end{center}
\caption{XRD intensity for SC sample is plotted against calculated intensity. Uranium site occupancy is fixed to 1.}
\label{XRD_SC}

\end{figure}

We discuss the details of X-ray diffraction data taken on UTe$_2$ samples.
In Figures \ref{XRD_NS}, \ref{XRD_NS_U0962} and \ref{XRD_SC}, observed X-ray intensity $I_{obs}$ is plotted against the calculated intensity $I_{calc}$. Here, only the data with relatively low intensity are shown. If the analysis is correct, the data points would be on the ideal line $I_{obs} = I_{calc}$ as shown by the solid line. For the UTe$_2$ NS sample, observed data show strong deviation from the ideal line, as shown in Fig. \ref{XRD_NS}. Among the data points shown by red dots, strongly deviating data are highlighted by blue squares. The deviation is much larger than the experimental error. Numbers in parentheses are reflection indices for these points. All these indices are characterized by the strong scattering factors from both uranium and tellurium but cancelling each other resulting in rather weak intensities. Such reflections are sensitive to deviation of atomic displacement factors or site occupancy. While the former effect is significant only in the low scattering angles, the latter affects all angle region. The indices shown in Fig. \ref{XRD_NS} spread in wide angle region. It is therefore likely that the deviations are due to the deviation of site occupancy. In fact, as described in the main text, changing the uranium site occupancy drastically improves the fitting (Fig. \ref{fit_XRD}). As shown in Figure \ref{XRD_NS_U0962}, the blue squares fall on the ideal line after correcting the occupancy. For SC samples, the $I_{obs}$ agrees with $I_{calc}$ for full occupancy (Fig. \ref{XRD_SC}).


\section{Uranium deficiency in UTe}
Uranium monotelluride with the NaCl-type structure was prepared by Bridgman technique. Starting uranium and tellurium with the composition 1:1 were loaded in a tungsten crucible. The crucible was then welded in an arc-furnace with an Ar atmosphere. The crucible was heated in a rf furnace up to approximately 1800 $^\circ$C then slowly cooled down.

The sample was characterized by the single-crystal X-ray diffraction as described in the main text. The lattice parameter was determined as $a$ = 6.1660(3) $\rm\AA$, in agreement with the reported value\cite{Slovyanskikh1976}. The occupancy parameter of the uranium site was determined as 0.918 where the agreement factor $R_1$ = 0.0238 takes a minimum.

\section*{References}

\bibliography{UTe2}

\providecommand{\newblock}{}
\begin{thebibliography}{10}
\expandafter\ifx\csname url\endcsname\relax
  \def\url#1{{\tt #1}}\fi
\expandafter\ifx\csname urlprefix\endcsname\relax\def\urlprefix{URL }\fi
\providecommand{\eprint}[2][]{\url{#2}}

\bibitem{Ran2019}
Ran S, Eckberg C, Ding Q~P, Furukawa Y, Metz T, Saha S~R, Liu I~L, Zic M, Kim
  H, Paglione J and Butch N~P 2019 {\em Science (80-. ).\/} {\bf 365} 684--687
  ISSN 10959203

\bibitem{Haneveld1970}
Haneveld A~J and Jellinek F 1970 {\em J. Less-Common Met.\/} {\bf 21} 45--49
  ISSN 00225088

\bibitem{Beck1988}
Beck H~P and Dausch W 1988 {\em Zeitschrift f{\"{u}}r Naturforsch. B\/} {\bf
  43} 1547--1550 ISSN 1865-7117

\bibitem{Suski1972a}
Suski W, Gibi{\'{n}}ski T, Wojakowski A and Czopnik A 1972 {\em Phys. Status
  Solidi\/} {\bf 9} 653--658 ISSN 00318965

\bibitem{Shlyk1995a}
Shlyk L, Tro{\'{c}} R and Kaczorowski D 1995 {\em J. Magn. Magn. Mater.\/} {\bf
  140} 1435--1436

\bibitem{Shlyk1999}
Shlyk L and Troc R 1999 {\em Physica B: Condens. Matter\/} {\bf 262} 90--97
  ISSN 09214526

\bibitem{Suski1973}
Suski W 1973 {\em J. Solid State Chem.\/} {\bf 7} 385--399 ISSN 1095726X

\bibitem{Ikeda2006}
Ikeda S, Sakai H, Aoki D, Homma Y, Yamamoto E, Nakamura A, Shiokawa Y, Haga Y
  and {\=O}nuki Y 2006 {\em J. Phys. Soc. Jpn.\/} {\bf 75} 116--118 ISSN
  0031-9015

\bibitem{Fujimori2019}
Fujimori S~i, Kawasaki I, Takeda Y, Yamagami H, Nakamura A, Homma Y and Aoki D
  2019 {\em J. Phys. Soc. Jpn.\/} {\bf 88} 1--5 ISSN 13474073
  (\textit{Preprint} \eprint{1908.09418})

\bibitem{Miao2020}
Miao L, Liu S, Xu Y, Kotta E~C, Kang C~J, Ran S, Paglione J, Kotliar G, Butch
  N~P, Denlinger J~D and Wray L~A 2020 {\em Phys. Rev. Lett.\/} {\bf 124} 76401
  ISSN 10797114 (\textit{Preprint} \eprint{1911.10152})

\bibitem{Knafo2021}
Knafo W, Knebel G, Steffens P, Kaneko K, Rosuel A, Brison J~P, Flouquet J, Aoki
  D, Lapertot G and Raymond S 2021 {\em Phys. Rev. B\/} {\bf 104} 1--6 ISSN
  24699969 (\textit{Preprint} \eprint{2106.13087})

\bibitem{Duan2021}
Duan C, Baumbach R~E, Podlesnyak A, Deng Y, Moir C, Breindel A~J, Maple M~B and
  Dai P 2021   1--18 (\textit{Preprint} \eprint{2106.14424})

\bibitem{Raymond2021}
Raymond S, Knafo W, Knebel G, Kaneko K, Brison J~P, Flouquet J, Aoki D and
  Lapertot G 2021 {\em J. Phys. Soc. Jpn.\/}  1--11 (\textit{Preprint}
  \eprint{2107.13914v1})

\bibitem{Sheikin2001}
Sheikin I, Huxley A~D, Braithwaite D, Brison J~P, Watanabe S, Miyake K and
  Flouquet J 2001 {\em Phys. Rev. B\/} {\bf 64} 220503

\bibitem{Levy2005a}
L{\'e}vy F, Sheikin I, Grenier B and Huxley A~D 2005 {\em Science (80-. ).\/}
  {\bf 309} 1343--1346 ISSN 0036-8075

\bibitem{Huy2008}
Huy N~T, de~Nijs D~E, Huang Y~K and de~Visser A 2008 {\em Phys. Rev. Lett.\/}
  {\bf 100} 077002 ISSN 0031-9007

\bibitem{Aoki2009a}
Aoki D, {D Matsuda} T, Taufour V, Hassinger E, Knebel G and Flouquet J 2009
  {\em J. Phys. Soc. Jpn.\/} {\bf 78} 113709 ISSN 0031-9015 (\textit{Preprint}
  \eprint{0910.1157})

\bibitem{Hayes2021}
Hayes I~M, Wei D~S, Metz T, Zhang J, Eo Y~S, Ran S, Saha S~R, Collini J, Butch
  N~P, Agterberg D~F, Kapitulnik A and Paglione J 2021 {\em Science (80-. ).\/}
  {\bf 373} 797--801 ISSN 10959203

\bibitem{Thomas2021}
Thomas S~M, Stevens C, Santos F~B, Fender S~S, Bauer E~D, Ronning F, Thompson
  J~D, Huxley A and Rosa P~F~S 2021 {\em Phys. Rev. B\/} {\bf 104} 224501 ISSN
  2469-9950 (\textit{Preprint} \eprint{2103.09194})

\bibitem{Ran2019a}
Ran S, Liu I~L, Eo Y~S, Campbell D~J, Neves P~M, Fuhrman W~T, Saha S~R, Eckberg
  C, Kim H, Graf D, Balakirev F, Singleton J, Paglione J and Butch N~P 2019
  {\em Nat. Phys.\/} {\bf 15} 1250--1254 ISSN 17452481 (\textit{Preprint}
  \eprint{1905.04343})

\bibitem{Braithwaite2019}
Braithwaite D, Vali{\v{s}}ka M, Knebel G, Lapertot G, Brison J~P, Pourret A,
  Zhitomirsky M~E, Flouquet J, Honda F and Aoki D 2019 {\em Commun. Phys.\/}
  {\bf 2} ISSN 23993650

\bibitem{Li2021}
Li D, Nakamura A, Honda F, Sato Y~J, Homma Y, Shimizu Y, Ishizuka J, Yanase Y,
  Knebel G, Flouquet J and Aoki D 2021 {\em J. Phys. Soc. Jpn.\/} {\bf 90} 1--5
  ISSN 13474073 (\textit{Preprint} \eprint{2105.08593})

\bibitem{Cairns2020}
Cairns L~P, Stevens C~R, O'Neill C~D and Huxley A 2020 {\em J. Phys. Condens.
  Matter\/} {\bf 32} ISSN 1361648X

\bibitem{Rosa2021}
Rosa P~F~S, Weiland A, Fender S~S, Scott B~L, Ronning F, Thompson J~D, Bauer
  E~D and Thomas S~M 2021  (\textit{Preprint} \eprint{2110.06200})

\bibitem{Sheldrick2015}
Sheldrick G~M 2015 {\em Acta Crystallogr. Sect. C Struct. Chem.\/} {\bf 71}
  3--8 ISSN 2053-2296

\bibitem{Momma2013}
Momma K, Ikeda T, Belik A~A and Izumi F 2013 {\em Powder Diffr.\/} {\bf 28}
  184--193 ISSN 08857156

\bibitem{Boehme1992}
Boehme D~R, Nichols M~C, Snyder R~L and Matheis D~P 1992 {\em J. Alloys
  Compd.\/} {\bf 179} 37--59 ISSN 09258388

\bibitem{KleinHaneveld1964}
{Klein Haneveld} A and Jellinek F 1964 {\em J. Inorg. Nucl. Chem.\/} {\bf 26}
  1127--1128 ISSN 00221902

\bibitem{Ran2021a}
Ran S, Liu I~l, Saha S~R, Saraf P, Paglione J and Butch N~P 2021 {\em J. Vis.
  Exp.\/}  1--9 ISSN 1940-087X

\bibitem{Aoki2019b}
Aoki D, Nakamura A, Honda F, Li D, Homma Y, Shimizu Y, Sato Y~J, Knebel G,
  Brison J~P, Pourret A, Braithwaite D, Lapertot G, Niu Q, Vali{\v{s}}ka M,
  Harima H and Flouquet J 2019 {\em J. Phys. Soc. Jpn.\/} {\bf 88} 043702 ISSN
  0031-9015 (\textit{Preprint} \eprint{1903.02410})

\bibitem{Aoki2021}
Aoki D, Brison J~P, Flouquet J, Ishida K, Knebel G, Tokunaga Y and Yanase Y
  2021   1--47 (\textit{Preprint} \eprint{2110.10451})

\bibitem{Willa2021}
Willa K, Hardy F, Aoki D, Li D, Wiecki P, Lapertot G and Meingast C 2021 {\em
  Phys. Rev. B\/} {\bf 104} 205107 ISSN 2469-9950 (\textit{Preprint}
  \eprint{2107.02706})

\bibitem{Coqblin1969}
Coqblin B and Schrieffer J~R 1969 {\em Phys. Rev.\/} {\bf 185} 847--853 ISSN
  0031-899X

\bibitem{Slovyanskikh1976}
Slovyanskikh V, Rozanov I and Gracheva N 1976 {\em Russ. J. Inorg. Chem.\/}
  {\bf 21} 1383--1386

\end{thebibliography}
\bibliographystyle{iopart-num}

\end{document}